\documentclass{aa}
\usepackage{graphicx}
\usepackage{txfonts}
\usepackage{hyperref}
\newcommand{\Teff}{$T_{\mathrm{eff}}$}          
\newcommand{\logg}{$\log g$}
\newcommand{\Msun}{$M_\sun$}

\begin{document}

\title{New white dwarf envelope models and diffusion}
\subtitle{Application to DQ white dwarfs}
   \author{D. Koester
          \inst{1}
          \and
          S.O. Kepler\inst{2}
          \and
          A.W. Irwin\inst{3}
          }
     \institute{Institut f\"ur Theoretische
     Physik und Astrophysik, Universit\"at Kiel, 24098 Kiel,
     Germany
     \and Instituto de Fisica, Universidade Federal do Rio Grande do Sul, 
     91501-900 Porto-Alegre, RS, Brazil
     \and Department of Physics and Astronomy, University of Victoria, PO Box
      1700 STN CSC, Victoria, BC V8W 2Y2, Canada 
     }
   \date{}

 \abstract
   {Recent studies of the atmospheres of carbon-rich (DQ) white dwarfs
     have demonstrated the existence of two different populations that are
     distinguished by the temperature range, but more importantly, by
     the extremely high masses of the hotter group. The classical DQ
     below 10\,000~K are well understood as the result of dredge-up
     of carbon by the expanding helium convection zone. The high-mass
     group poses several problems regarding their origin and also an
     unexpected correlation of effective temperature with mass.}
   {We propose to study the envelopes of these objects to
     determine the total hydrogen and helium masses as possible clues
     to their evolution.}
   {We developed new codes for envelope integration and diffusive
     equilibrium that are adapted to the unusual chemical composition,
     which is not necessarily dominated by hydrogen and helium.}
   {Using the new results for the atmospheric parameters, in
     particular, the masses obtained using {\it Gaia} parallaxes, we
     confirm that the narrow sequence of carbon abundances with
     \Teff\ in the cool classical DQ is indeed caused by an almost
     constant helium to total mass fraction, as found in earlier
     studies. This mass fraction is smaller than predicted by stellar
     evolution calculations. For the warm DQ above 10\,000~K, which
     are thought to originate from double white dwarf mergers, we
     obtain extremely low hydrogen and helium masses. The correlation
     of mass with \Teff\ remains unexplained, but another possible
     correlation of helium layer masses with \Teff\ as well as the
     gravitational redshifts casts doubt on the reality of both and
     suggests possible shortcomings of current models.}
   {}

   \keywords{Stars: white dwarfs -- Stars: carbon -- Stars: evolution
     -- Convection -- Diffusion}

   \maketitle

\section{Introduction}
White dwarfs with the major features from carbon in the form of
molecular Swan bands or atomic carbon lines are called DQ white
dwarfs. Most of them, at least below effective temperatures of about
10\,000~K, have helium-dominated atmospheres, although this element
cannot be seen directly. Several recent studies
\citep{Koester.Kepler19, Coutu.Dufour.ea19, Blouin.Dufour19} have
analyzed large samples provided by the Sloan Digital Sky Survey
\citep[SDSS,][]{York.Adelman.ea00, Abolfathi.Aguado.ea18,
  Kepler.Pelisoli.ea19}, and even more importantly, the {\it Gaia} DR2
parallax measurements \citep{GaiaCollaboration.Brown.ea18}, to further
our understanding of these unusual objects. It is now firmly
established that the DQ form two distinct populations. The first are
the classical cool DQ at cooler \Teff\ with traces of carbon, which
are visible as molecular Swan bands of C$_2$, and masses very similar
to those of the main groups of hydrogen-rich (DA) and helium-rich (DB,
DC, and DZ) white dwarfs. The other group are the so-called warm and
hot DQ between about 10\,000 and 24\,000~K, which are characterized by
\ion{C}{i} and at the hot end also \ion{C}{ii} atomic lines. These
objects differ from the other group mainly by their high masses above
0.9~\Msun, but also by their chemical composition. The atmospheres of
several of the warm and all of the hot DQ are dominated by carbon,
with significant contributions of hydrogen also visible. Helium is
possibly the major element in the cool region between 10\,000 and
14\,000~K, but no spectral features are seen, and the exact C-to-He
ratio is difficult to determine.

There is general agreement that the origin of the cool DQ is dredge-up
of carbon by the growing He convection zone as the star cools
\citep{Koester.Weidemann.ea82, Pelletier.Fontaine.ea86}. The warm DQ
may be descendants of the hot DQ at higher \Teff, but the high masses
of both groups seem to exclude a normal evolution from higher
\Teff\ helium-rich progenitors such as DB or DO white dwarfs. The only
currently discussed solution seems to be that they form as a result of
a merger of two white dwarfs \cite{Dunlap.Clemens15}. Not only are the
masses unusually high: the recent studies \citet{Koester.Kepler19} and
\citet{Coutu.Dufour.ea19} found a strong correlation of masses
increasing with \Teff\ between 10\,000 and 16\,000~K, which rules out
a normal cooling evolution for these objects and is not understood.

In this study we try to look below the visible surface of cool and
warm DQ by calculating the envelope stratification, starting from
atmosphere models with the observed parameters. This allows us to
estimate the total mass of the lighter elements, hydrogen and
helium.  For the classical DQ this has been done before
\cite{Pelletier.Fontaine.ea86, Dufour.Bergeron.ea05}, but it can now be
repeated with improved knowledge about the masses of the objects, and
with improved input physics in the envelope calculations. One of the aims
is to determine the mass of the helium layer, which can be compared to
predictions from evolutionary calculations. For the warm DQ such a
study is not yet available to our knowledge. The hope is that
determination of the total hydrogen and helium masses can shed light
on the outcome of mergers, and perhaps help to resolve open problems.

The envelope code we used in past projects 
  \citep[e.g.,][]{Koester09} was written for stars whose outer layers
are dominated by hydrogen and helium. Because this may not be true for all
DQ, we have written a more general code and also updated the equation of
state and opacity calculations. We also used recent results in
the literature to improve the physics for the diffusion equilibrium
calculations. The next section describes the methods and input
physics, which are then applied to the case of cool and warm DQ.
 
\section{Envelope equations and input physics} The structure of the
stellar envelope is governed by the four stellar structure equations
for mass conservation, hydrodynamic equilibrium, energy transport, and
energy generation. Using the mass $m$ inside radius $r$ as independent
variable, these can be written as
\begin{eqnarray} \frac{dr}{dm} & = &
  \frac{1}{4\pi r^2 \rho} \\
  \frac{dP}{dm} & = & -\frac{G m}{4 \pi r^4}  \label{eqp}\\
  \frac{d\ln T}{d\ln P} & = & \frac{3 P \kappa}{64 \pi \sigma T^4 G}\,
  \frac{l}{m} \mbox{\qquad (radiative)}\\
  & = & \nabla_{conv} \mbox{\qquad (convective)}\nonumber\\
  \frac{dl}{dm} & = & \mbox{const} = \frac{L}{M}
,\end{eqnarray}
with pressure $P$, temperature $T$, mass density $\rho$, luminosity
$l(m)$, $G$ the gravitation constant, $\sigma$ the radiation constant,
and $\kappa$ the absorption coefficient. The convective gradient
$\nabla_{conv}$ is calculated with the mixing-length approximation in
the ML2 version \citep{Tassoul.Fontaine.ea90}, assuming a mixing
length of 1.25 pressure scale heights.  $M$ and $L$ are the total mass
and luminosity of the model. The last equation for the energy
generation assumes zero nuclear energy generation, and that the
gravitational energy generation is roughly proportional to the
mass. In practice, this is almost equivalent to assuming $l =
\mbox{const}$ because of the very low mass of the envelope.  We make
the equations dimensionless using stellar mass $M$ and radius $R$ as
normalization as well as
\begin{eqnarray}
  P_c &=& \frac{G M^2}{R^4} \\
  T_c &=& \left(\frac{L}{4\pi \sigma R^2}\right)^{1/4} \\
  \rho_c &=& \frac{3M}{4\pi R^3}
.\end{eqnarray}
The independent variable is changed to \begin{equation} x = \ln(1 -
  m/M) .\end{equation} The three resulting equations are integrated
downward into the star to a level $q = \log(1-m/M)$ (typically -2.0,
i.e., $m/M = 0.99$ or $(M-m)/M = 0.01$ of the stellar mass in the outer
envelope), which is specified as a parameter. The algorithm we used is a
Runge-Kutta method as described in \cite{Press.Teukolsky.ea92}.

\subsection{Boundary conditions}
The boundary conditions are determined at some layer of a stellar
atmosphere model. This gives the pressure $P_0$, density $\rho_0$, and
temperature $T_0$, and also the effective temperature \Teff\ and
surface gravity \logg. From the latter two and a mass-radius relation
we can obtain the total mass $M$, radius $R$, and luminosity $L$. The
starting value for $x$  is obtained by integrating eq. \ref{eqp} from the
surface ($P = 0, m = M, r = R$) downward to the boundary layer.

\subsection{Equation of state  and opacities \label{EOS}}
The development of the current code was motivated by the study of the
outer envelopes of carbon atmosphere white dwarfs of spectral types DQ
(classical, warm, and hot). For these we cannot assume that either
hydrogen or helium or a mixture of these are the dominating
constituents of the atmospheres and envelopes. In several warm DQs and
most hot DQs, carbon seems to be the most abundant element. In our
previous envelope calculations we used the equation of state (EOS) of
\citet{Saumon.Chabrier.ea95}, which considers only H/He mixtures. We
therefore completely rewrote the code and included as a further option
for the EOS the ``FreeEOS'' \citep{Cassisi.Salaris.ea03, Irwin12},
which can include the 20 most important elements up to atomic number
28 (Ni).

Similarly, for the opacities, we have three options. As in previous
versions of our envelope code we can use the OPAL Rosseland opacity
tables for ten H/He mixtures \citep{Iglesias.Rogers96}. Alternatives
are the OPAL opacity tables for H/He mixtures with (normal) metal
content $Z = 0$, but enhanced carbon and oxygen content between $0$
and $1$, or the Los Alamos OP tables for 20 single elements from H to
Ni \citep{Colgan.Kilcrease.ea16}. In all these cases the conductive
opacities are from \cite{Potekhin.Baiko.ea99,
  Potekhin.Pons.ea15}.\footnote{ \url{www.ioffe.ru/astro/conduct/}}

\subsection{Diffusion \label{diff}}
In convection zones (cvz) the turbulent velocities are many orders of
magnitude higher than typical diffusion velocities, and the matter
will remain completely homogeneously mixed. However, below the bottom
of this zone, or below any overshooting zone if included,
gravitational settling will change the element stratification. For the
envelopes of white dwarfs the diffusion timescales are always shorter
than the evolution timescales, and we therfore assume that diffusion
equilibrium is achieved. This means that abundance gradients are
formed that balance the gravitational settling, and diffusion fluxes
are zero. At the lower boundary of the convection zone some
overshooting will necessarily be present. We assume that this, as well
as diffusion caused by any abundance gradient, will ensure a smooth
transition for the element abundances.

Following the arguments in \citet{Pelletier.Fontaine.ea86} and
\citet{Paxton.Schwab.ea18}, we neglect thermal effects, and all terms
involving collisions with electrons. The diffusion equations for a
multicomponent plasma then take the form \citep{Burgers69}
\begin{equation}
  d_i = \frac{d p_i}{d r} + n_i A_i m_u g - n_i Z_i e E
  =  \sum_{i \ne k} K_{ik} (w_i - w_k)
  .\end{equation}
Here $d_i$ is the driving force on ion $i$, provided by the partial
pressure $p_i$, the local gravity $g,$ and electric field $E$. The
atomic mass and charge units are $m_u$ and $e$, atomic mass $A_i$ and
average charge $Z_i$, number density $n_i$. The relative velocity
between ions $i$ and $k$ is $w_i - w_k$, and $K_{ik}$ are the
resistance coefficients, which are related in a simple way to the
usual diffusion coefficients $D_{ik}$ \citep{Pelletier.Fontaine.ea86}.

In the case of diffusion equilibrium the relative velocities are zero,
and the system of equations simplifies to a set of equations for the
local abundance gradients,
\begin{equation}
  \frac{d p_i}{d r} =  -n_i A_i m_u g + n_i Z_i e E \label{eq10}
.\end{equation}
For the electrons the gravitational term can be neglected:
\begin{equation}
  -n_e e E = \frac{d p_e}{d r} = \sum_j Z_j \frac{d p_j}{d r}
,\end{equation}
where the first equality comes from eq.~\ref{eq10}, and the second
from the condition of electrical neutrality. This result serves to
eliminate the electric field and leaves $N$ linear equations
for $N$ ions.

These equations are derived and transformed under the assumption of a
classical ideal gas EOS for electrons and ions. In the
envelopes of white dwarfs two complications arise: the electrons may
be partially degenerate, and Coulomb interactions in the plasma may
not be negligible. We took the degeneracy into account by
replacing
\begin{equation}
  p_e = n_e kT \mbox{\quad with \quad} p_e = f n_e kT
\end{equation}
with the Boltzmann constant $k$ and $f$ determined from the Fermi-Dirac
functions for partial degeneracy of the electrons.

For the consideration of non-ideal effects of the ions we follow
\cite{Beznogov.Yakovlev13}. They replace
\begin{equation}
  \frac{d p_i}{d r} \rightarrow n_i \frac{d
    \mu_i}{d r} =  \frac{d p_i}{d r} + 
     n_i \frac{d \mu_i^c}{d r}
,\end{equation}
where now $p_i$ is the ideal gas part as before, $\mu_i$ is the
chemical potential of the ion, and $\mu_i^c$ is the Coulomb interaction
contribution. For the latter, they derive
\begin{equation}
  \frac{d \mu_i^c}{d r} = -0.3 \frac{Z_i^{5/3}
    e^2}{r_e} \frac{d \ln n_e}{d r}
,\end{equation}
with
\begin{equation}
  r_e = \left(\frac{3}{4\pi n_e}\right)^{1/3}
.\end{equation}
At each depth in the envelope the system of equations for $d
p_i/d r $ is solved and integrated alongside the three
  structure equation. This defines the abundance profiles $\epsilon_i(m)$.
  
As can be seen in the diffusion equations, we need the average charges
of all ions $Z_i$. In previous versions we used the approximation from
\citet{Paquette.Pelletier.ea86*b} (corrected for a missing factor of
$\rho^{1/3}$), which is based on the rather crude pressure ionization
model of \citet{Fontaine.Michaud79}. This model also assumes a trace
element in a uniform background, which is not appropriate in our case,
as shown below. New options are a Thomas-Fermi mean-ionization model
\citep{Stanton.Murillo16, More85}, which is based originally on
\citet{Feynman.Metropolis.ea49}.  This method elegantly takes into
account the multicomponent nature of the plasma and is our preferred
choice now.  A third option are tables of mean charges for 20 elements
from H to Ni, provided together with the Los Alamos opacity tables
\citep{Colgan.Kilcrease.ea16}.

\section{Application to DQ white dwarfs}
While individual results show some differences between the two recent
studies of \citet{Koester.Kepler19} and \citet{Coutu.Dufour.ea19}, the
important results agree and are discussed in the following. For these
applications we used for the cool DQ the EOS of
\citet{Saumon.Chabrier.ea95}, if the abundances by number $\epsilon(H)
+ \epsilon(He) > 0.999$, and FreeEOS elsewhere. For the warm DQ, we
used FreeEOS everywhere. The opacity was calculated from the OPAL
tables for H, He, C, O mixtures. The average charges were determined
with the Thomas-Fermi mean-ionization model.

\subsection{Cool DQ}
For the classical cool DQ with Swan bands, the main result is the
appearance of a clearly defined sequence of objects from
\textasciitilde 10\,000~K, [C/He] = -4.0 down to 5500~K and [C/He] =
\hbox{-7.0} (the notation [C/He] is used as abbreviation for
abundances of $\log(n(\mathrm{C})/n(\mathrm{He}))$. Above this
sequence other objects are scattered whose carbon abundances are up to
1~dex higher, possibly forming a second sequence. While these general
results were already demonstrated in earlier work
\citep{Dufour.Bergeron.ea05, Koester.Knist06}, the {\it Gaia}
parallaxes now confirm that the DQ on the low abundance sequence
(henceforth called only ``the sequence'') have (almost) normal white
dwarf masses.

Because the standard explanation for these cool DQ is dredge-up of
carbon from the underlying He/C transition zone
\citep{Koester.Weidemann.ea82, Pelletier.Fontaine.ea86}, the most
natural parameter for the formation of the sequence is the thickness
or total mass of the He layer. The latter study found that a
relatively thin helium layer with $q$(He) in the range $-3.5$ to
$-4.0$ provided the best fit to the observed carbon
abundances. Improved calculations shown in
\citet{Dufour.Bergeron.ea05} indicate a slightly higher range of
$-2.5$ to $-3.0$, but it is still smaller than the predicted $\ge$-2.0
from evolutionary calculations.

While the cited papers used evolutionary calculations for white dwarfs
including the effects of diffusion, in our current study we follow a
complementary approach. At least the outer envelope at these low
effective temperatures is certainly in diffusive equilibrium,
therefore we integrate the envelope equations from the outside, with
atmosphere models using observed abundances as starting point. The
structure is followed inward through the convection zone (present in
all DQ), where the matter is homogeneously mixed, and below, where the
abundances change as described in the previous section. The
integration is typically stopped at $q=-2.0$, or when the abundance
fraction (by mass) of helium is below $10^{-4}$, whatever is deeper in
the star.

Rather than using every observed DQ, we decided to describe the
sequence with eight representative models. We divided the range from 9750
to 5750~K into 500~K wide intervals and determined average values
for \Teff, \logg, and [C/He], excluding all objects with [C/He]
greater than the average by more than 0.4 (to exclude the objects
that lie clearly above the sequence). The parameters of the resulting
model sequence are given in Table\ref{modseq}. 

\begin{table}
  \caption{Parameters \Teff, \logg, [C/He], and standard deviation of
    the abundance distribution for eight models describing the observed
    sequence of cool DQ. The first column gives the number of objects
    in the interval. }
\label{modseq} 
\centering         
\begin{tabular}{rrrrr}
\hline\hline
\noalign{\smallskip}
  No.   & \Teff [K] & \logg   &  [C/He]  & $\sigma$[C/He]\\
  \hline
  \noalign{\smallskip}
    4& 9335 & 7.961 & -4.304 & 0.161 \\
   14& 8942 & 7.954 & -4.511 & 0.096 \\
   46& 8469 & 7.889 & -4.852 & 0.076 \\
   48& 8009 & 7.954 & -5.207 & 0.182 \\
   37& 7503 & 7.979 & -5.605 & 0.179 \\
   18& 7053 & 8.071 & -5.923 & 0.134 \\
   17& 6490 & 7.923 & -6.542 & 0.166 \\
    3& 5900 & 7.856 & -6.973 & 0.136 \\
\hline\\
\end{tabular}
\end{table}

The \logg\ averages in the intervals are close enough to the overall
average 7.95, and we used this value for all models, varying only
\Teff\ and [C/He].  Atmospheric models were calculated with these
parameters. Pressure and temperature at a deep layer (Rosseland $\tau
= 90$), together with radius, mass, and luminosity obtained from
\Teff\ and \logg\ from the Montreal mass-radius
relation \footnote{\url{www.astro.umontreal.ca/~bergeron/CoolingModels}}
were used as starting values for the envelope equations. Throughout
the cvz, the abundances were held fixed at the atmospheric values;
below the bottom they were changed according to the conditions of
diffusive equilibrium. The resulting fractional masses in the cvz as
well as the total fractional helium mass for different assumptions are
given in Table~\ref{modcvz}. The different options are no overshoot,
one pressure scale height overshoot below the formal bottom of the
cvz, and a calculation without overshoot, but with the non-ideal term
in the diffusion equations switched off.

The 19 objects excluded from the determination of the model sequence
because of higher C abundances have an average \logg\ = 7.94, that is,
their masses are identical with the cool sequence. We calculated
additional atmosphere and envelope models for three parameter sets at
the low and top end of the sequence and one in the middle with [C/He]
increased by 1~dex compared to the standard. We also calculated three
models with [C/He] decreased until the carbon features became
undetectable, defined as a jump $<1\%$ for the strong band head at
5163~\AA\ (after convolving for the SDSS resolution). These six
envelope models are presented in Table \ref{dqhighlow} and are
compared with the standard sequence.

\begin{table}
  \caption{Fractional mass in the convection zone q(cvz) = $ \log
    M_{cvz}/M $. $\eta$ is the electron degeneracy parameter at the
    bottom of the cvz, $\Gamma_e = e^2/r_e kT $ the parameter
    measuring the Coulomb interaction compared to the thermal
    energy. q$_0$(He), q$_1$(He), and q$_2$(He) are the total He mass
    fraction in the envelope without overshoot, assuming one pressure
    scale height overshoot, and with the non-ideal terms in the
    diffusion equations switched off. The last two rows give the
    average and standard deviation (width of the distribution) for the
    helium mass fractions.
\label{modcvz} }
\centering         
\begin{tabular}{rrrrrrr}
\hline\hline
\noalign{\smallskip}
\Teff   & q(cvz) & $\eta$  &  $\Gamma_e$  & q$_0$(He) &
q$_1$(He) & q$_2$(He) \\
\hline
\noalign{\smallskip}
 5900 & -4.76 & 17.4 & 1.8 & -3.55 & -3.16 & -2.69\\
 6490 & -4.85 & 11.5 & 1.3 & -3.52 & -3.14 & -2.61\\
 7053 & -4.90 &  7.6 & 0.9 & -3.46 & -3.09 & -2.54\\
 7503 & -4.90 &  5.9 & 0.7 & -3.36 & -3.00 & -2.45\\
 8009 & -4.91 &  4.7 & 0.6 & -3.28 & -2.92 & -2.40\\
 8469 & -4.93 &  3.8 & 0.5 & -3.25 & -2.90 & -2.40\\
 8942 & -4.98 &  3.1 & 0.4 & -3.27 & -2.91 & -2.45\\
 9335 & -5.02 &  2.6 & 0.4 & -3.29 & -2.93 & -2.48\\
  \hline
 average &    &      &     & -3.37 & -3.00 & -2.50\\
  sd  &       &      &     &  0.13 &  0.11 &  0.10\\
\hline\\
\end{tabular}
\end{table}

\begin{table}
  \caption{Envelope calculations for three representative models above
    the sequence with [C/He] enhanced by 1~dex, and with [C/He]
    decreased until the carbon features become undetectable in the
    SDSS spectra. These are compared with the standard sequence
    (middle of the subsections). 
\label{dqhighlow} }
\centering
\begin{tabular}{rrrr}
\hline\hline
\noalign{\smallskip}
  \Teff & [C/He] & q(cvz) & q(He)\\
\hline
\noalign{\smallskip}
5900 & -7.97 & -4.52 & -3.26 \\
5900 & -6.97 & -4.75 & -3.55 \\
5900 & -5.97 & -5.00 & -3.85 \\
\hline
7503 & -6.61 & -4.77 & -3.17 \\
7503 & -5.61 & -4.91 & -3.35 \\
7503 & -4.61 & -5.06 & -3.58 \\
\hline
9335 & -4.80 & -4.92 & -3.10 \\
9335 & -4.30 & -5.03 & -3.29 \\
9335 & -3.30 & -5.29 & -3.74 \\
\hline
\end{tabular}
\end{table}

\begin{figure}
\centering
\includegraphics[width=0.60\textwidth, angle=-90]{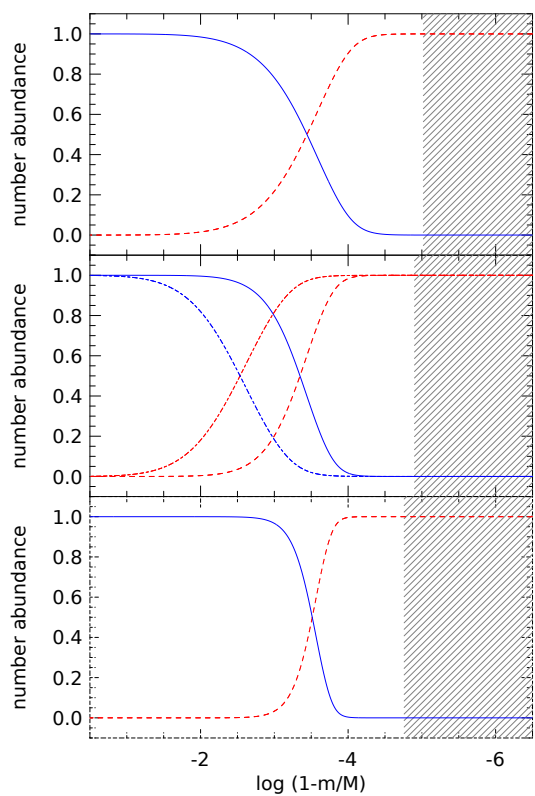}
\caption{Number abundances for He and C in the lower part of the
  convection zone and below the bottom. The three panels show from top
  to bottom the models for \Teff\ = 9335, 7503, and 5900~K.  The curves
  give abundances for He (long red dashes, declining toward the interior
  on the left side) and C (continuous blue, declining toward the
  surface). The short-dashed curves in the middle panel result from
  switching off the non-ideal interactions in the diffusion equations,
  which results in a higher He mass by about a factor 10. The hatched
  area is the cvz.
\label{figcool}}
\end{figure}

With our standard assumptions for the sequence (no overshoot, with
non-ideal terms) we obtain helium layer masses that are slightly
higher than in \cite{Pelletier.Fontaine.ea86}, but lower than the more
recent results in \cite{Dufour.Bergeron.ea05} or
\cite{Coutu.Dufour.ea19}. We note, however, that these results depend
strongly on the input physics used. When we include overshooting, the
He mass increases by \textasciitilde 0.4~dex, and when we neglect the
non-ideal term in the diffusion equations, they increase by
\textasciitilde 0.9~dex (without overshooting!). Another ingredient
that is very important for the depth of the cvz is the conductive
opacity. Several scale heights above the bottom, the temperature
gradient is already dominated by the conductive opacity, but it is
still high enough to enforce convection. Thus the bottom of the cvz is
determined by the decrease in conductive opacity.

The helium masses show a small systematic decrease in the lower part
of the model sequence. Fig.~12 in \citet{Coutu.Dufour.ea19} also shows
a slight deviation of the observed objects from a theoretical sequence
with constant q(He), although the direction is different from our
case: the coolest objects have higher He masses than the hotter ones,
which contradicts our result. Nevertheless, the small change of
\textasciitilde 0.3~dex from warm to cool models is much smaller than
the 2.5~dex span of carbon abundances observed at the surface. The
explanation of the sequence as due to stars with very similar He
masses is therefore probably correct, and small apparent trends are
due to remaining imperfections of our models, especially at the cool
end.

That the deepest convection zones (at 5900~K) lead to the lowest He
masses disagrees with naive expectations. The main reason for this is
the evolution of the average charge of carbon ions along the sequence
from hot to cool \Teff: at the bottom of the cvz, the temperature
decreases, while the density increases, leading to a decrease of the
average charge from 5.0 to 4.1. The gradients in the C/He transition
zone become steeper and the total He mass decreases in spite of a
slight increase in cvz mass. This change in average C charge is also
the main reason for the huge decrease of carbon abundances along the
sequence to the cool end. This is demonstrated in Fig.~\ref{figcool},
which shows the number abundances of He and C in the envelopes as a
function of effective temperature. While the total He content remains
approximately constant, the gradients become much steeper with the
decreasing average carbon charge, and the tail reached by the cvz
accordingly has much lower C abundance. A similar effect has been
noted before, and its importance was realized by
\citet{Pelletier.Fontaine.ea86}.

The effect of including the Coulomb interaction in the diffusion
equations is demonstrated in the middle panel of Fig.~\ref{figcool},
which shows the C/He abundance structure for the 7503~K model through
the lower part of the cvz and in the region below. The interaction
term leads to a steeper decline of the He abundance and thus to a
lower total He mass fraction.

One reason for the small change in apparent He masses at low
\Teff\ could be an unseen hydrogen content in the DQ, which is not
considered in our cool models.  This has been discussed by
\citet{Coutu.Dufour.ea19}, who concluded that the addition of [H/He] =
-3.0 in the models does not significantly change the results. We find
that even [H/He] = -4.0 would easily be detected and have used such a
model grid for some test calculations.  At the low \Teff\ end
(5900~K), the change of the model corresponds to a change in optical
slope of \textasciitilde 50K hotter, and [C/He] higher by 0.1~dex.
The flux at 4800~\AA\ is about 0.05 mag higher, leading to a smaller
radius estimate and an increase in log g by 0.05~dex. At higher
\Teff\ , the changes are smaller and completely negligible at the high
\Teff\ end. The same is true for the cvz depths and He masses
(Table~\ref{modcvzh}), which show only very minor differences to the
models without hydrogen.

The additional models in Table~\ref{dqhighlow} show that the cool DQ
sequence spans only a rather narrow range of helium masses. A very
small decrease leads to the few scattered objects above the sequence,
while about 0.2~dex more renders the carbon features undetectable, and
the objects would appear as featureless DC white dwarfs. This is of
course influenced by observational bias. The white dwarf L\,97-3 shows
a completely featureless optical spectrum and was originally
classified as a DC, until \cite{Koester.Weidemann.ea82} found strong
\ion{C}{i} lines in the ultraviolet region and determined an abundance
of [C/He] = -6.0. Using the same methods as for the DQ, we find a
fractional helium mass $q$(He) = -2.99, only 0.3~dex higher than would
make the object a DQ in the visible region as well.  This suggests a
narrow distribution of He masses from stellar evolution, with the
classical DQ formed by the low-mass tail and the majority of objects
evolving as DC. Whether a continuous distribution can lead to such a
sharply defined sequence, or if some special feature in this function
is needed remains to be tested by stellar evolution calculations and
population synthesis simulations. In this context, it will also be
important to confirm (or refute) the apparent slight difference of
0.05~\Msun\ between the average DQ and DC masses
\citep{Koester.Kepler19,Coutu.Dufour.ea19}.

\begin{table}
  \caption{Same as Table~\ref{modcvz}, but for models with
    [H/He] = -4.0. 
\label{modcvzh} }
\centering         
\begin{tabular}{rrrrrrr}
\hline\hline
\noalign{\smallskip}
  \Teff   & q(cvz) & $\eta$  &  $\Gamma_e$  & q(He)0 & q(He)1 & q(He)2 \\
\hline
\noalign{\smallskip}
 5900 & -4.92 & 13.4 & 1.5 & -3.67 & -3.28 & -2.77\\
 6490 & -4.96 &  9.3 & 1.1 & -3.56 & -3.18 & -2.62\\
 7053 & -4.97 &  6.7 & 0.8 & -3.46 & -3.09 & -2.54\\
 7503 & -4.95 &  5.4 & 0.7 & -3.35 & -3.00 & -2.44\\
 8009 & -4.96 &  4.3 & 0.6 & -3.29 & -2.93 & -2.40\\
 8469 & -4.97 &  3.5 & 0.5 & -3.26 & -2.91 & -2.41\\
 8942 & -5.00 &  2.9 & 0.4 & -3.28 & -2.92 & -2.45\\
 9335 & -5.04 &  2.5 & 0.4 & -3.30 & -2.94 & -2.50\\
  \hline
  ave &       &      &     & -3.38 & -3.01 & -2.50\\
  sd  &       &      &     &  0.16 &  0.15 &  0.14\\
\hline\\
\end{tabular}
\end{table}

\begin{table*}
  \caption{Atmospheric parameters \Teff, \logg\ (cgs units), [H/C],
    and  [He/C] for 26 warm DQ white dwarfs. q(cvz) is the fractional mass
    $\log(M_{cvz}/M)$ in the convection zone, q(H) and q(He) are the
    logarithmic fractional masses of hydrogen and helium. The two last
    rows show the average and standard deviation for the H and He
    masses, assuming that the upper limits are really present. We also
    list the individual masses and spectroscopic radial velocities
    ($V_\mathrm{r}$), which are not corrected for gravitational redshifts.
\label{dqwarm} }
\centering         
\begin{tabular}{rrrrrrrrrr}
\hline\hline
\noalign{\smallskip}
  SDSSJ &\Teff/K &\logg  & $M$/\Msun & $V_\mathrm{r}$ [km/s] &[H/C]
  &[He/C] &q(cvz) &q(H) &q(He)\\
\hline
\noalign{\smallskip}
001908.63+184706.0& 11000& 8.686& 1.01& 110&$<$-2.30& $<$1.70&  -7.57&$<$-11.82& $<$-6.44\\
023633.74+250348.9& 14611& 8.777& 1.07& 130&   -2.00& $<$1.00&  -8.30&   -11.61& $<$-7.15\\
080708.49+194950.0& 14593& 8.847& 1.11& 220&   -1.80& $<$0.00&  -8.62&   -11.21& $<$-8.05\\
085626.94+451337.0&  9353& 8.509& 0.91&    &$<$-0.60&    2.90&  -6.75&$<$-10.46&    -5.46\\ 
085914.63+325712.2& 10300& 8.605& 0.97& 120&$<$-2.30&    2.20&  -7.20&$<$-11.92&    -5.99\\
091922.26+023604.4& 11500& 8.702& 1.02& 170&   -1.70&    2.20&  -7.50&   -11.61&    -6.22\\
093638.04+060709.6& 12166& 8.816& 1.09& 100&$<$-1.80& $<$2.00&  -7.87&$<$-11.89& $<$-6.62\\  
095837.00+585303.0& 15444& 8.951& 1.16& 100&   -0.70& $<$0.00&  -9.12&   -10.63& $<$-8.54\\
104906.61+165923.7& 13590& 8.995& 1.19& 130&   -1.25& $<$1.25&  -8.60&   -11.37& $<$-7.50\\
105817.66+284609.3& 10500& 8.696& 1.02& 120&$<$-2.30&    2.20&  -7.42&$<$-12.14&    -6.23\\ 
110058.04+175807.0& 12631& 8.756& 1.05& 100&   -2.80& $<$1.20&  -8.01&   -11.30& $<$-6.90\\
114006.29+073529.9& 11400& 8.738& 1.04& 120&   -1.00&    2.20&  -7.58&   -10.99&    -6.33\\
114059.88+182401.9& 10500& 8.620& 0.98& 120&$<$-2.30&    2.00&  -7.30&$<$-11.83&    -6.12\\
114851.68$-$012612.7&10500&8.610& 0.97& 130&$<$-2.30&    2.20&  -7.22&$<$-11.94&    -6.00\\
120331.77+645059.6& 12400& 8.700& 1.02& 100&   -2.00& $<$1.00&  -7.91&   -11.24& $<$-6.87\\
121510.66+470010.3& 13940& 8.957& 1.16& 100&   -1.00& $<$1.00&  -8.64&   -10.96& $<$-7.59\\
133151.38+372754.8& 16741& 9.028& 1.16& 240&   -1.40&$<$-0.40&  -9.07&   -11.20& $<$-8.82\\
133221.56+235502.2& 15131& 8.779& 1.07& 100&   -1.30& $<$0.60&  -8.53&   -10.87& $<$-7.53\\
133940.50+503613.5& 11680& 8.621& 0.98& 160&$<$-1.80&    2.00&  -7.40&-$<$11.41&    -6.12\\
134124.28+034628.7& 13978& 8.834& 1.10& 170&   -2.20& $<$1.20&  -8.32&   -10.80& $<$-7.16\\
143437.82+225859.5& 15750& 8.828& 1.09& 130&   -1.70& $<$0.00&  -8.70&   -11.19& $<$-8.07\\
143534.01+531815.1& 15658& 8.900& 1.13& 100&   -1.25& $<$0.75&  -8.78&   -11.16& $<$-7.71\\
144854.80+051903.5& 15966& 8.943& 1.16& 100&   -1.10&$<$-0.30&  -8.95&   -10.78& $<$-8.60\\
162205.12+184956.7& 16693& 9.129& 1.16& 120&   -1.60& $<$0.10&  -9.37&   -11.80& $<$-8.73\\
162236.13+300454.5& 16131& 8.934& 1.15& 100&   -1.40&$<$-0.30&  -8.88&   -11.01& $<$-8.52\\
172856.22+555822.8& 14772& 8.869& 1.12& 100&   -1.50&$<$-0.30&  -8.64&   -10.88& $<$-8.33\\
\hline
average  &               & 8.801& 1.07& 128&      &                  &  &-11.31&    -7.22\\
sd       &               & 0.152& 0.08&  38&      &                  &  &  0.46&     1.02\\
\hline\\
\end{tabular}
\end{table*}

\begin{figure}
\centering
\includegraphics[width=0.34\textwidth, angle=-90]{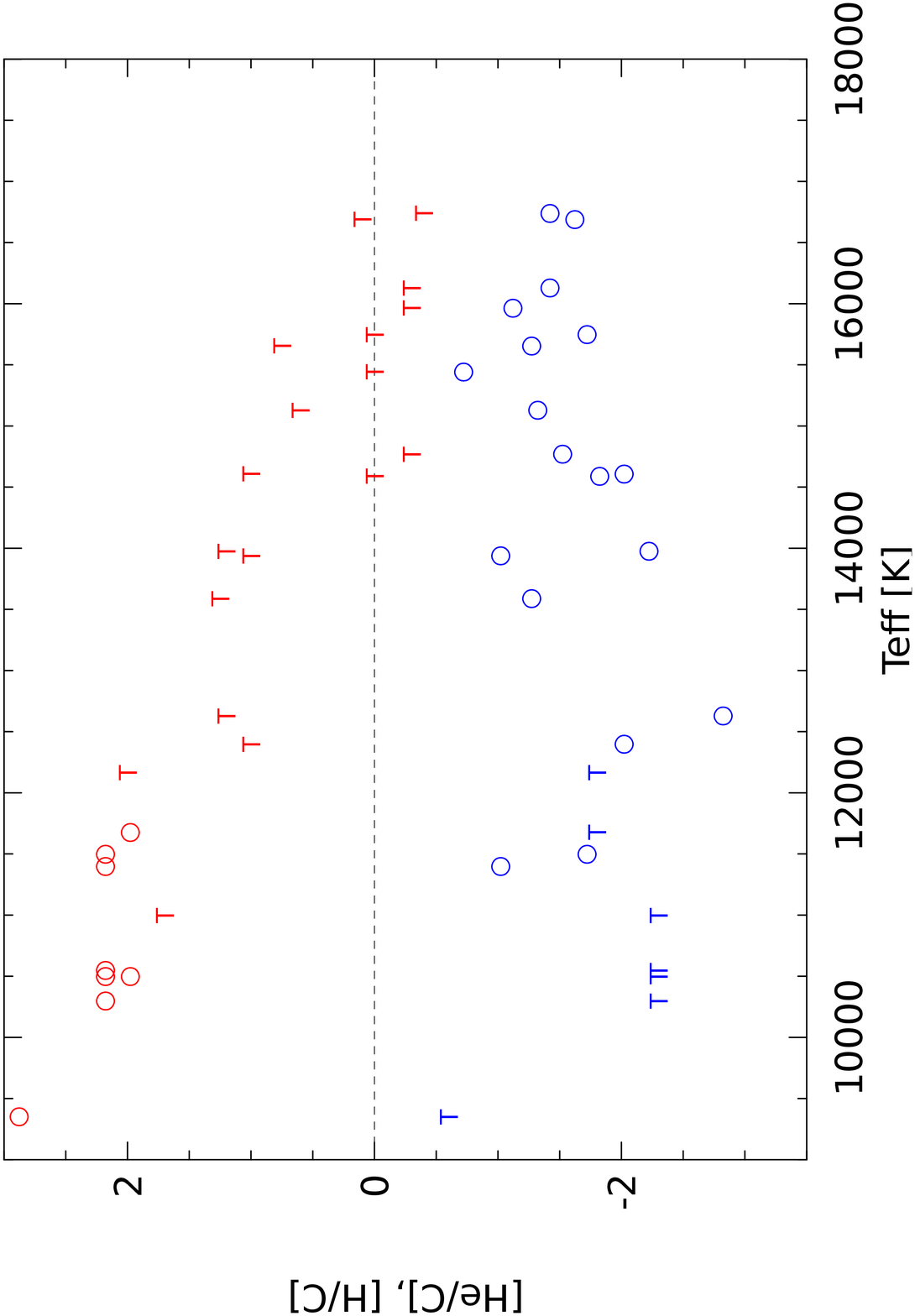}
\caption{Atmospheric [H/C] and [He/C] abundance ratios and upper
  limits for warm DQ white dwarfs. Upper part (red) [He/C], and lower
  part (blue) [H/C]. The horizontal dashed line is the zero-point
  [He/C] = [H/C] = 0.
\label{wdqabund}}
\end{figure}

\subsection{Warm DQ}
Warm DQ is the designation given to a group of carbon-rich white
dwarfs at temperatures between \textasciitilde 10\,000-17\,000~K,
whose spectra are dominated by atomic lines of neutral carbon
(\ion{C}{i}). The new $Gaia$ parallaxes have shown for the first time
that all these stars have extremely high surface gravities above
$8.50$, corresponding to masses $>0.9$~\Msun \citep{Koester.Kepler19,
  Coutu.Dufour.ea19}. In their analysis, \citet{Koester.Kepler19}
noted that many of these objects showed hydrogen Balmer lines, but
they kept the abundance fixed at [H/He]=-3.5, referring further
analysis to the current study.

Our sample consists of 26 warm DQ with parameters \Teff, \logg, [C/He]
given in the earlier study. These were the starting point for our
determination of a more accurate H content. We realized, however, that
a changed H content and closer inspection of the spectra demanded some
changes in the parameters. In the range of 9300 - 11000~K, weak
\ion{C}{i} lines, which had a negligible influence on the $\chi^2$
fitting, sometimes demanded a slightly higher \Teff \ because they
were too weak in the models. We therefore increased the temperature
(and \logg\ as demanded by the parallax) until the fit was more
satisfactory. We note that this is not a final high-accuracy analysis
because our interest is only to obtain reasonably accurate starting
models for the envelope integration.

\begin{figure}
\centering
\includegraphics[width=0.60\textwidth, angle=-90]{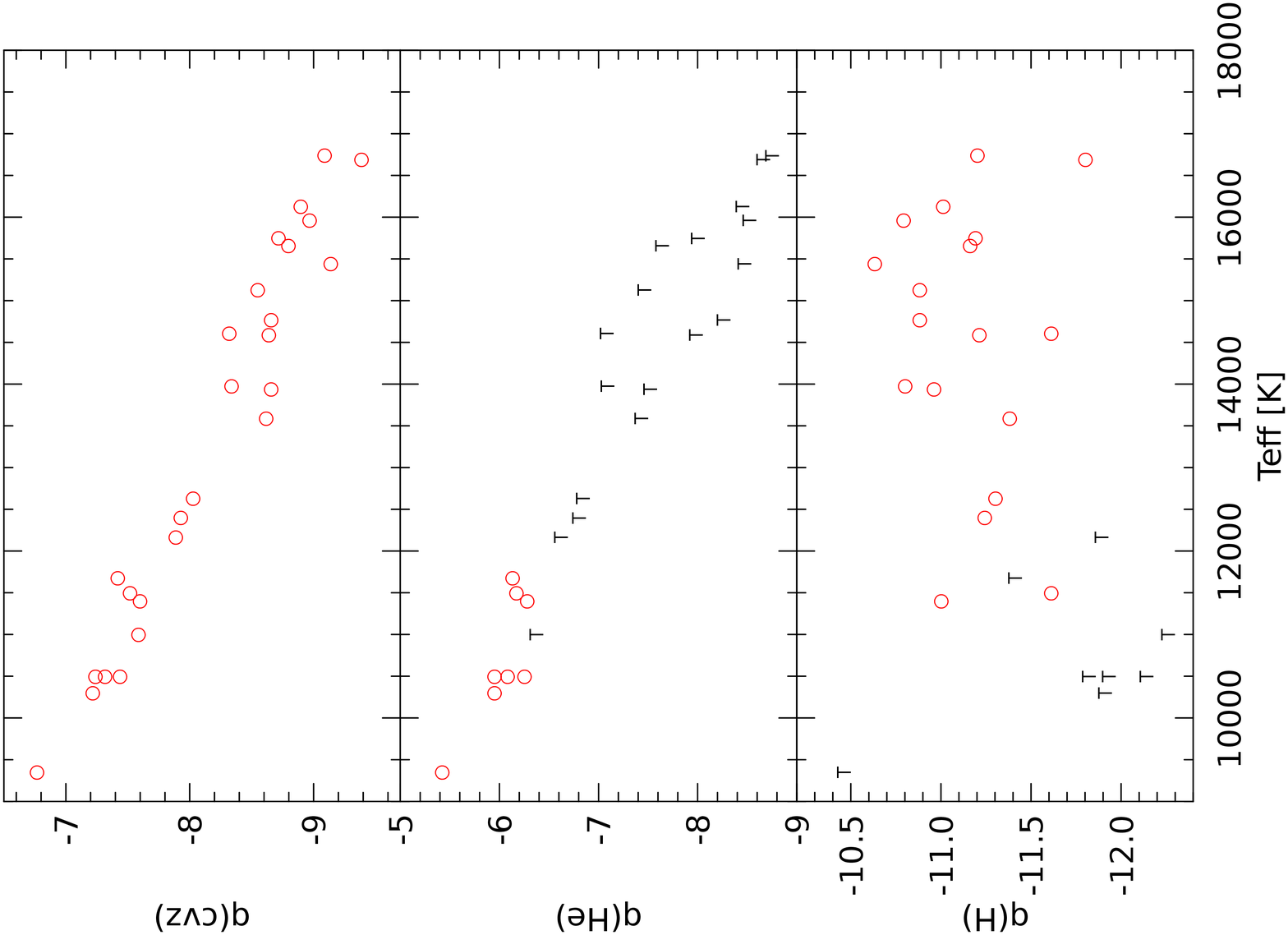}
\caption{Logarithmic fractional convection zone masses, helium, and
  hydrogen masses (red circles) and upper limits (black) for warm DQ.
\label{wdqenv}}
\end{figure}

\begin{figure}
\centering
\includegraphics[width=0.34\textwidth, angle=-90]{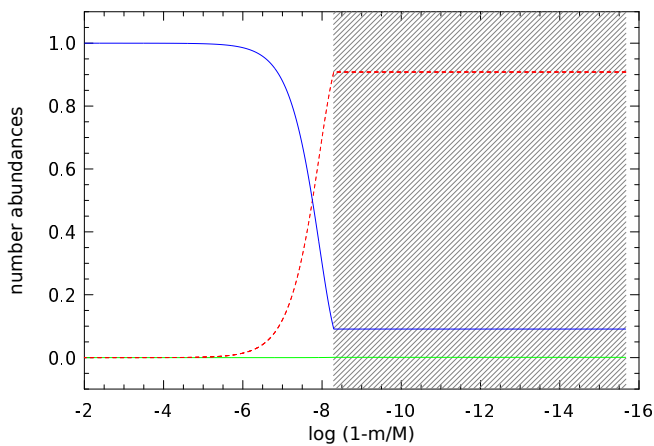}
\caption{Abundance structure of H (green, lowest curve), He (red,
  dashed), and C (blue, continuous) in SDSSJ0236+2503. On this
  linear scale the H abundance cannot be distinguished from zero. The
  hatched area is the convection zone.
\label{wdqstruct}}
\end{figure}

\begin{figure}
\centering
\includegraphics[width=0.60\textwidth, angle=-90]{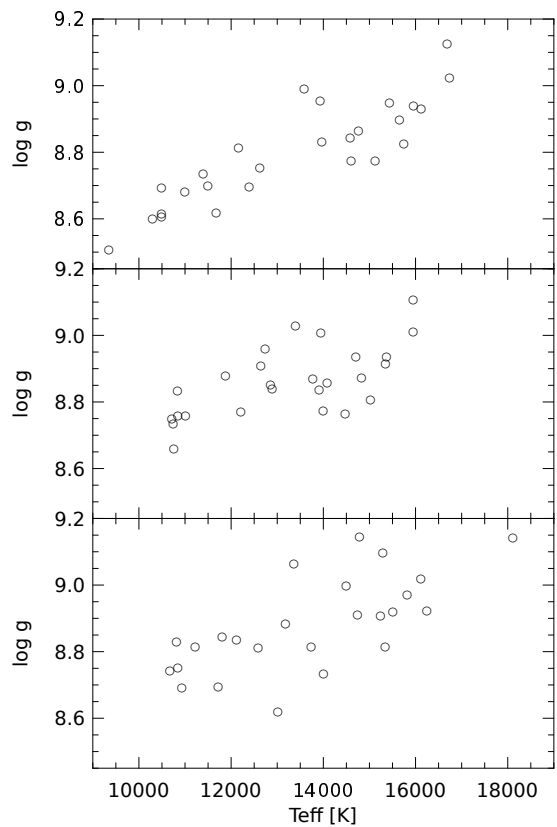}
\caption{Surface gravities as function of \Teff. Top panel: Standard
  results, with [He/C], \Teff, and \logg\ as variables. Middle panel:
  [H/C] = -2.0 and [He/C] = -1.50 fixed, only photometry is used to
  determine \Teff\ and \logg.  Bottom panel: Results after iterating
  between \logg\ from photometry and \Teff\ from spectra, with [H/C]
  and [He/C] fixed as above.
\label{wdqnohe}}
\end{figure}

A more serious problem is presented by the helium-to-carbon ratio at
temperatures $>$11500~K, with [C/He] $>-2.0$ The [C/He] ratio obtained
by the $\chi^2$ fitting often produces a small but visible \ion{He}{i}
line 5877~\AA, which is not observed in any of the warm DQ we
analyzed. This is also true when we use the parameters of the common
objects from \citet{Coutu.Dufour.ea19} with our own models. When we
increased the [C/He] ratio until this line became compatible with
observations, in most cases we realized that the complete \ion{C}{i}
spectrum does not change with further increase, that is, the He
abundance can only be regarded as an upper limit. This is unexpected
because the $\chi^2$ fitting clearly leads to a solution with
significant He contribution, but it might be explained with the
extremely unsatisfactory status of the atomic data for neutral carbon
in the optical range \citep{Koester.Kepler19}. For the lower
\Teff\ range and [He/C] $>2.0$ there is still a small dependency of
the carbon features on the abundance ratio, and we kept these results
as measurement; some doubt remains whether these might in reality be
just upper limits as well.

Carbon is always observed, whereas helium is not directly observed in
any of the objects. We therefore preferred to change our abundance
notation to using carbon as standard, and measuring He and H relative
to C. The final parameters used for the 26 objects are given in
Table~\ref{dqwarm}, which also has the results for the depth of the
convection zone and total H and He masses (which in the latter case
are mostly upper limits). Figure~\ref{wdqabund} shows the atmospheric
H and He abundances and upper limits for the whole range of
temperatures.  The cvz depths and H and He mass fractions are
displayed in Fig.~\ref{wdqenv}. A typical example
(SDSSJ023633.74+250348.9) for the abundance structure in the envelope
is provided in Fig.~\ref{wdqstruct}.

Figure~\ref{wdqenv} shows that the mass in the cvz decreases with
increasing \Teff. If the cooler and warmer objects were evolutionary
related, we would expect the atmospheric He abundance to decrease with
cooling because the cvz becomes more massive. The opposite seems
to happen - the He abundances are apparently higher at the low
\Teff\ end. We recall, however, that most if not all of the He
abundances are upper limits, and the trend in the
abundances at least partly reflects the increasing visibility of the
He lines.

\subsection{The puzzle of surface gravities} The studies of
\citet{Koester.Kepler19} and \citet{Coutu.Dufour.ea19} both reported a
clear correlation of the surface gravity with temperature from
10000-17000~K that reached very high values at the hot end
(Fig.\ref{wdqnohe}, top panel). Both studies failed to identify
shortcomings in the models (completely independent codes).  This does
not rule out that shortcomings can be found eventually because
some arguments cast doubt on this \Teff\ - \logg\ correlation.

First of all, we note that while the surface gravities in
Table~\ref{dqwarm} extend over more than a factor of four, the spread
in masses is only 0.28~\Msun\ or about 26\%, with a strong clustering
between 1.05 and 1.16~\Msun. Because of the shape of the mass-radius
relation at the high-mass end, a small change in mass leads to a much
larger change of surface gravity.  Another argument is given by the
radial velocities. All objects show high redshifts $\ge 100$~km/s,
with an average of 128 and a standard deviation of 38~km/s, and no
correlation with \Teff. We assume that the standard deviation is
caused by the space motion, which is not unrealistic because the
median of the transverse velocity is \textasciitilde 50~km/s
\citep{Coutu.Dufour.ea19}. The average gravitational redshift is then
128~km/s, which at 13\,000~K corresponds to a surface gravity of 8.98
and a mass of 1.17~\Msun.  This is higher than the average mass from
\logg\ (1.07~\Msun), which might in part be caused by the strongly
nonlinear relation between mass and gravitational redshift. This
number might also indicate the result of the merger of two white
dwarfs with the most likely masses of \textasciitilde 0.58~\Msun.

Our result that most of the [He/C] results should be regarded as upper
limits, as well the recent detection of a DAQ near \Teff\ = 11\,000~K
with very unusual abundances (Hollands et al. 2020, in preparation),
encouraged us to make a rather radical experiment. We used a model
grid with [He/C] fixed at -1.50, which means that its influence on the
spectra is negligible, and [H/C] fixed at -2.0, a typically observed
value. The results for the surface gravities are shown in
Fig.~\ref{wdqnohe}. One uses only photometry, and the other iterates
between photometry for \logg\ and spectroscopy for \Teff\ . The standard results are included as well. The scatter in the two lower panels is
larger than in the top, most likely because only two
parameters are free in the fitting procedure. Nevertheless, the trend
toward higher masses at higher temperatures is still apparent and
seems to be a robust result, independent of the details of the
analysis.

\section{Discussion and conclusions}
The total helium masses we find, even with overshooting included, are
lower than predicted from stellar evolution. In their study of the
full evolution from the main sequence to white dwarfs, including
element diffusion, \citet{Althaus.Panei.ea09} and
\citet{MillerBertolami.Althaus06} found (log) He mass fractions of
-1.99 for a white dwarf with 0.584~\Msun, with a range of -1.88 to
-2.59 for white dwarf masses between 0.542 and
0.741~\Msun. \cite{Romero.Kepler.ea13} calculated the complete
evolution leading to ZZ Ceti variables, that is, DA with outer
hydrogen envelope. They also presented numbers for the resulting He
layer mass, which is -1.62 for a DA mass of
0.593~\Msun. \cite{Romero.Corsico.ea12} argued that the He mass might
decrease by a factor of four at most, when they enforced a large
number of thermal pulses on the giant branch. However, in that case,
the white dwarf mass increases to \textasciitilde 0.7~\Msun.

Such He masses will lead to a pure He atmosphere and a white dwarf
classified as DC (featureless spectrum) below 10\,000~K effective
temperature, or a DZ, if traces of metals other than C are present.
\cite{Kepler.Pelisoli.ea19} listed 2445 DC+DZ in the SDSS Data Release
14, and only 524 DQ. While this is certainly not a statistically valid
sample, the analysis of \cite{Giammichele.Bergeron.ea12} of the white
dwarfs within 20~pc lists about twice as many DZ+DC as DQ. The cool
DQ are obviously not the majority among cool He-rich white dwarfs, but
a significant fraction. It is therefore very likely that the relation
between white dwarf (or progenitor) mass and final He layer mass is
not strictly a one-to-one relation, but has some statistical element,
perhaps with the exact number of thermal pulses experienced. The cool
DQ then would originate from the lower tail of resulting helium
masses.

The situation is far more complex for the warm DQs.  Whether the He
and H masses are real determinations or only upper limits, they are
clearly much lower than would be expected from normal single-star
evolution. As there is also no known single star progenitor population
with this mass distribution, the currently favored explanation is that
these massive objects are the results of the merger of two white
dwarfs, as first proposed by \cite{Dunlap.Clemens15} for the hot
DQ. \cite{Coutu.Dufour.ea19} extended this to the warm DQ, which are
considered to be the cooled-down descendants. For the hot DQ the
argument rests on the discrepancy between the transversal velocities,
which are characteristic for an older population, and the younger
cooling age.  \cite{Cheng.Cummings.ea19} also concluded based on
kinematic arguments that \textasciitilde 20\% of all white dwarfs in
the range 0.8-1.3~\Msun\ are the product of double white dwarf
mergers. Their objects are, however, mostly hotter than 16\,000~K, and
the argument is less convincing for the warm DQ because of the longer
cooling ages even for single stars. This leaves the high masses as the
single argument.  Many authors have studied the merger of two white
dwarfs, but the emphasis usually is on the origin of supernova Ia
events, and thus on the merger of high-mass white dwarfs with a total
mass above the Chandrasekhar mass. \cite{Sato.Nakasato.ea15} also
reported lower mass mergers, with individual stars of 0.5 and
0.6~\Msun, where the outcome is not an explosion but a massive white
dwarf. All these mergers apparently go through a phase of spiral-in
with a massive disk, and it seems plausible that such an evolution can
remove almost the complete outer H and He layers, leaving such small
amounts as we find in the warm DQ. To our knowledge, no detailed
computations of the outer layers exist, however.

A major remaining problem is the apparent trend between \logg\ and
\Teff, which seems to indicate that there is no evolutionary
connection between the very massive objects near 16000~K and the less
massive ones below 13000~K (which are still more massive than the cool
DQ). Another argument is the increase in He abundance and also He
masses with decreasing \Teff, although this has to assume that the
atmospheric He abundances are real and not upper
limits. If this is confirmed, it would be highly unlikely that these
two puzzles are unrelated, possibly suggesting a major shortcoming of
current model atmospheres.

\cite{Coutu.Dufour.ea19} proposed explaining the
mass-\Teff\ correlation as a pile-up of objects due to a slow-down of
the cooling because of the interior crystallization and its dependence
on stellar mass. Their Fig.~17 is somewhat deceiving, however, because
the shown isochrones correspond to differences in
\emph{\textup{logarithmic age}}. Using the Montreal single-star
evolution calculations, we find that the cooling age of a DQ with
1.05~\Msun\ is 0.38~Gy from 16\,000-13\,000~K, but 0.92~Gy from
13\,000 to 10\,000~Gy, yet there are far fewer objects in the second
interval than in the first. It is also unknown whether the isochrones
have any relevance if the sample of warm DQ does not constitute a
cooling sequence. To compare with theoretical models, calculations are
required that include binary interaction, as in
\citet{Istrate.Marchant.ea16}.

The final question is where the cooled-down warm DQ are. They cannot
avoid this fate, but there are very few if any DQ below 9500~K with
masses above 0.8~Msun. \cite{Koester.Kepler19} have suggested,
although with rather questionable arguments, that the so-called
peculiar DQ (DQpec) with shifted Swan bands might be descendants of
the warm DQ. \cite{Blouin.Dufour19} have refuted this and found normal
masses for the DQpec as well. In their interpretation, a DQ turns into
a DQpec when the photospheric density exceeds 0.15~g/cm$^3$, which
occurs when they cool down to below 6500~K (depending also on carbon
abundance). Six DQpec above this line are suggested to be probably
magnetic. The results strongly depend on the temperature
determinations: if they were higher, the surface gravities would also
be higher. While the theoretical models of \citet{Blouin.Dufour19} and
\citet{Blouin.Dufour.ea19} represent the best currently possible
effort, they still have to use uncertain approximations. The
calculations by \cite{Kowalski10} of the density shift of the Swan
bands show a rather large effect; \cite{Blouin.Dufour.ea19}
recalibrated the shift by an empirical factor of eight so that it
agreed with the observed shifts. We completely agree with the
statement by \cite{Blouin.Dufour19} that ``more efforts on both the
observational and theoretical fronts are needed to clarify the nature
of these objects''.

On the observational fronts, spectra with high resolution and high S/N
of warm DQ in the range 11\,000-13\,000~K could help to detect tiny He
lines. Perhaps it would be even easier to perform UV observations of
the region 1900-3500~\AA, where large differences between helium-rich
and helium-poor models are expected.

\begin{acknowledgement}
This work was financed in part by the Coordena\c{c}\~ao de
Aperfei\c{c}oamento de Pessoal de N\'{\i}vel Superior - Brasil (CAPES)
- Finance Code 001, Conselho Nacional de Desenvolvimento
Cient\'{\i}fico e Tecnol\'ogico - Brasil (CNPq), and Funda\c{c}\~ao de
Amparo \`a Pesquisa do Rio Grande do Sul (FAPERGS) - Brasil.  This
research has made use of public data from the Sloan Digital Sky Survey
and the {\it Gaia} Mission. 
\end{acknowledgement}


\begin{thebibliography}{37}
\expandafter\ifx\csname natexlab\endcsname\relax\def\natexlab#1{#1}\fi

\bibitem[{{Abolfathi} {et~al.}(2018){Abolfathi}, {Aguado}, {Aguilar},
    {Allende Prieto}, {Almeida}, {Ananna}, {Anders}, {Anderson},
    {Andrews}, {Anguiano}, \& et~al.}]{Abolfathi.Aguado.ea18}
  {Abolfathi}, B., {Aguado}, D.~S., {Aguilar}, G., {et~al.} 2018,
  \apjs, 235, 42

\bibitem[{{Althaus} {et~al.}(2009){Althaus}, {Panei}, {Miller
      Bertolami}, {Garc{\'{\i}}a-Berro}, {C{\'o}rsico}, {Romero},
    {Kepler}, \& {Rohrmann}}]{Althaus.Panei.ea09} {Althaus}, L.~G.,
  {Panei}, J.~A., {Miller Bertolami}, M.~M., {et~al.} 2009, \apj, 704,
  1605

\bibitem[{{Beznogov} \& {Yakovlev}(2013)}]{Beznogov.Yakovlev13}
  {Beznogov}, M.~V. \& {Yakovlev}, D.~G. 2013, \prl, 111, 161101

\bibitem[{{Blouin} \& {Dufour}(2019)}]{Blouin.Dufour19} {Blouin},
  S. \& {Dufour}, P. 2019, \mnras, 490, 4166

\bibitem[{{Blouin} {et~al.}(2019){Blouin}, {Dufour}, {Thibeault}, \&
    {Allard}}]{Blouin.Dufour.ea19} {Blouin}, S., {Dufour}, P.,
  {Thibeault}, C., \& {Allard}, N.~F. 2019, \apj, 878, 63

\bibitem[{{Burgers}(1969)}]{Burgers69} {Burgers}, J.~M. 1969, {Flow
  Equations for Composite Gases. Academic Press, New York}

\bibitem[{{Cassisi} {et~al.}(2003){Cassisi}, {Salaris}, \&
    {Irwin}}]{Cassisi.Salaris.ea03} {Cassisi}, S., {Salaris}, M., \&
  {Irwin}, A.~W. 2003, \apj, 588, 862

\bibitem[{{Cheng} {et~al.}(2019){Cheng}, {Cummings}, {M{\'e}nard}, \&
    {Toonen}}]{Cheng.Cummings.ea19} {Cheng}, S., {Cummings}, J.~D.,
  {M{\'e}nard}, B., \& {Toonen}, S. 2019, arXiv e-prints,
  arXiv:1910.09558

\bibitem[{{Colgan} {et~al.}(2016){Colgan}, {Kilcrease}, {Magee},
    {Sherrill}, {Abdallah}, {Hakel}, {Fontes}, {Guzik}, \&
    {Mussack}}]{Colgan.Kilcrease.ea16} {Colgan}, J., {Kilcrease},
  D.~P., {Magee}, N.~H., {et~al.} 2016, \apj, 817, 116

\bibitem[{{Coutu} {et~al.}(2019){Coutu}, {Dufour}, {Bergeron},
    {Blouin}, {Loranger}, {Allard}, \& {Dunlap}}]{Coutu.Dufour.ea19}
  {Coutu}, S., {Dufour}, P., {Bergeron}, P., {et~al.} 2019, \apj, 885,
  74

\bibitem[{{Dufour} {et~al.}(2005){Dufour}, {Bergeron}, \&
    {Fontaine}}]{Dufour.Bergeron.ea05} {Dufour}, P., {Bergeron}, P.,
  \& {Fontaine}, G. 2005, \apj, 627, 404

\bibitem[{{Dunlap} \& {Clemens}(2015)}]{Dunlap.Clemens15} {Dunlap},
  B.~H. \& {Clemens}, J.~C. 2015, in Astronomical Society of the
  Pacific Conference Series, Vol. 493, 19th European Workshop on White
  Dwarfs, ed. P.~{Dufour}, P.~{Bergeron}, \& G.~{Fontaine}, 547

\bibitem[{Feynman {et~al.}(1949)Feynman, Metropolis, \&
    Teller}]{Feynman.Metropolis.ea49} Feynman, R.~P., Metropolis, N.,
  \& Teller, E. 1949, Phys. Rev., 75, 1561

\bibitem[{{Fontaine} \& {Michaud}(1979)}]{Fontaine.Michaud79}
  {Fontaine}, G. \& {Michaud}, G. 1979, \apj, 231, 826

\bibitem[{{Gaia Collaboration} {et~al.}(2018){Gaia Collaboration},
    {Brown}, {Vallenari}, {Prusti}, {de Bruijne}, {Babusiaux},
    {Bailer-Jones}, {Biermann}, {Evans}, {Eyer}, \&
    et~al.}]{GaiaCollaboration.Brown.ea18} {Gaia Collaboration},
  {Brown}, A.~G.~A., {Vallenari}, A., {et~al.} 2018, \aap, 616, A1

\bibitem[{{Giammichele} {et~al.}(2012){Giammichele}, {Bergeron}, \&
    {Dufour}}]{Giammichele.Bergeron.ea12} {Giammichele}, N.,
  {Bergeron}, P., \& {Dufour}, P. 2012, \apjs, 199, 29

\bibitem[{{Iglesias} \& {Rogers}(1996)}]{Iglesias.Rogers96}
  {Iglesias}, C.~A. \& {Rogers}, F.~J. 1996, \apj, 464, 943

\bibitem[{{Irwin}(2012)}]{Irwin12} {Irwin}, A.~W. 2012, ascl:1211.002

\bibitem[{{Istrate} {et~al.}(2016){Istrate}, {Marchant}, {Tauris},
    {Langer}, {Stancliffe}, \& {Grassitelli}}]{Istrate.Marchant.ea16}
  {Istrate}, A.~G., {Marchant}, P., {Tauris}, T.~M., {et~al.} 2016,
  \aap, 595, A35

\bibitem[{{Kepler} {et~al.}(2019){Kepler}, {Pelisoli}, {Koester},
    {Reindl}, {Geier}, {Romero}, {Ourique}, {Oliveira}, \&
    {Amaral}}]{Kepler.Pelisoli.ea19} {Kepler}, S.~O., {Pelisoli}, I.,
  {Koester}, D., {et~al.} 2019, \mnras, 486, 2169

\bibitem[{{Koester}(2009)}]{Koester09} {Koester}, D. 2009, \aap, 498,
  517

\bibitem[{{Koester} \& {Kepler}(2019)}]{Koester.Kepler19} {Koester},
  D. \& {Kepler}, S.~O. 2019, \aap, 628, A102

\bibitem[{{Koester} \& {Knist}(2006)}]{Koester.Knist06} {Koester},
  D. \& {Knist}, S. 2006, \aap, 454, 951

\bibitem[{{Koester} {et~al.}(1982){Koester}, {Weidemann}, \&
    {Zeidler}}]{Koester.Weidemann.ea82} {Koester}, D., {Weidemann},
  V., \& {Zeidler}, E.-M. 1982, \aap, 116, 147

\bibitem[{{Kowalski}(2010)}]{Kowalski10} {Kowalski}, P.~M. 2010, \aap,
  519, L8

\bibitem[{{Miller Bertolami} \&
    {Althaus}(2006)}]{MillerBertolami.Althaus06} {Miller Bertolami},
  M.~M. \& {Althaus}, L.~G. 2006, \aap, 454, 845

\bibitem[{{More}(1985)}]{More85} {More}, R.~M. 1985, Advances in
  Atomic and Molecular Physics, 21, 305

\bibitem[{{Paquette} {et~al.}(1986){Paquette}, {Pelletier},
    {Fontaine}, \& {Michaud}}]{Paquette.Pelletier.ea86*b} {Paquette},
  C., {Pelletier}, C., {Fontaine}, G., \& {Michaud}, G. 1986, \apjs,
  61, 197

\bibitem[{{Paxton} {et~al.}(2018){Paxton}, {Schwab}, {Bauer},
    {Bildsten}, {Blinnikov}, {Duffell}, {Farmer}, {Goldberg},
    {Marchant}, {Sorokina}, {Thoul}, {Townsend}, \&
    {Timmes}}]{Paxton.Schwab.ea18} {Paxton}, B., {Schwab}, J.,
  {Bauer}, E.~B., {et~al.} 2018, \apjs, 234, 34

\bibitem[{{Pelletier} {et~al.}(1986){Pelletier}, {Fontaine},
    {Wesemael}, {Michaud}, \& {Wegner}}]{Pelletier.Fontaine.ea86}
  {Pelletier}, C., {Fontaine}, G., {Wesemael}, F., {Michaud}, G., \&
  {Wegner}, G.  1986, \apj, 307, 242

\bibitem[{{Potekhin} {et~al.}(1999){Potekhin}, {Baiko}, {Haensel}, \&
    {Yakovlev}}]{Potekhin.Baiko.ea99} {Potekhin}, A.~Y., {Baiko},
  D.~A., {Haensel}, P., \& {Yakovlev}, D.~G. 1999, \aap, 346, 345

\bibitem[{{Potekhin} {et~al.}(2015){Potekhin}, {Pons}, \&
    {Page}}]{Potekhin.Pons.ea15} {Potekhin}, A.~Y., {Pons}, J.~A., \&
  {Page}, D. 2015, \ssr, 191, 239

\bibitem[{{Press} {et~al.}(1992){Press}, {Teukolsky}, {Vetterling}, \&
    {Flannery}}]{Press.Teukolsky.ea92} {Press}, W.~H., {Teukolsky},
  S.~A., {Vetterling}, W.~T., \& {Flannery}, B.~P.  1992, Numerical
  recipes in FORTRAN. The art of scientific computing (Cambridge:
  University Press, 2nd ed.)

\bibitem[{{Romero} {et~al.}(2012){Romero}, {C{\'o}rsico}, {Althaus},
    {Kepler}, {Castanheira}, \& {Miller
      Bertolami}}]{Romero.Corsico.ea12} {Romero}, A.~D.,
  {C{\'o}rsico}, A.~H., {Althaus}, L.~G., {et~al.} 2012, \mnras, 420,
  1462

\bibitem[{{Romero} {et~al.}(2013){Romero}, {Kepler}, {C{\'o}rsico},
    {Althaus}, \& {Fraga}}]{Romero.Kepler.ea13} {Romero}, A.~D.,
  {Kepler}, S.~O., {C{\'o}rsico}, A.~H., {Althaus}, L.~G., \& {Fraga},
  L. 2013, \apj, 779, 58

\bibitem[{{Sato} {et~al.}(2015){Sato}, {Nakasato}, {Tanikawa},
    {Nomoto}, {Maeda}, \& {Hachisu}}]{Sato.Nakasato.ea15} {Sato}, Y.,
  {Nakasato}, N., {Tanikawa}, A., {et~al.} 2015, \apj, 807, 105

\bibitem[{{Saumon} {et~al.}(1995){Saumon}, {Chabrier}, \& {van
      Horn}}]{Saumon.Chabrier.ea95} {Saumon}, D., {Chabrier}, G., \&
  {van Horn}, H.~M. 1995, \apjs, 99, 713

\bibitem[{{Stanton} \& {Murillo}(2016)}]{Stanton.Murillo16} {Stanton},
  L.~G. \& {Murillo}, M.~S. 2016, \pre, 93, 043203

\bibitem[{{Tassoul} {et~al.}(1990){Tassoul}, {Fontaine}, \&
    {Winget}}]{Tassoul.Fontaine.ea90} {Tassoul}, M., {Fontaine}, G.,
  \& {Winget}, D.~E. 1990, \apjs, 72, 335

\bibitem[{{York} {et~al.}(2000){York}, {Adelman}, {Anderson},
    {Anderson}, {Annis}, {Bahcall}, {Bakken}, {Barkhouser}, {Bastian},
    {Berman}, {Boroski}, {Bracker}, {Briegel}, {Briggs}, {Brinkmann},
    {Brunner}, {Burles}, {Carey}, {Carr}, {Castander}, {Chen},
    {Colestock}, {Connolly}, {Crocker}, {Csabai}, {Czarapata},
    {Davis}, {Doi}, {Dombeck}, {Eisenstein}, {Ellman}, {Elms},
    {Evans}, {Fan}, {Federwitz}, {Fiscelli}, {Friedman}, {Frieman},
    {Fukugita}, {Gillespie}, {Gunn}, {Gurbani}, {de Haas}, {Haldeman},
    {Harris}, {Hayes}, {Heckman}, {Hennessy}, {Hindsley}, {Holm},
    {Holmgren}, {Huang}, {Hull}, {Husby}, {Ichikawa}, {Ichikawa},
    {Ivezi{\'c}}, {Kent}, {Kim}, {Kinney}, {Klaene}, {Kleinman},
    {Kleinman}, {Knapp}, {Korienek}, {Kron}, {Kunszt}, {Lamb}, {Lee},
    {Leger}, {Limmongkol}, {Lindenmeyer}, {Long}, {Loomis}, {Loveday},
    {Lucinio}, {Lupton}, {MacKinnon}, {Mannery}, {Mantsch}, {Margon},
    {McGehee}, {McKay}, {Meiksin}, {Merelli}, {Monet}, {Munn},
    {Narayanan}, {Nash}, {Neilsen}, {Neswold}, {Newberg}, {Nichol},
    {Nicinski}, {Nonino}, {Okada}, {Okamura}, {Ostriker}, {Owen},
    {Pauls}, {Peoples}, {Peterson}, {Petravick}, {Pier}, {Pope},
    {Pordes}, {Prosapio}, {Rechenmacher}, {Quinn}, {Richards},
    {Richmond}, {Rivetta}, {Rockosi}, {Ruthmansdorfer}, {Sandford},
    {Schlegel}, {Schneider}, {Sekiguchi}, {Sergey}, {Shimasaku},
    {Siegmund}, {Smee}, {Smith}, {Snedden}, {Stone}, {Stoughton},
    {Strauss}, {Stubbs}, {SubbaRao}, {Szalay}, {Szapudi}, {Szokoly},
    {Thakar}, {Tremonti}, {Tucker}, {Uomoto}, {Vanden Berk},
    {Vogeley}, {Waddell}, {Wang}, {Watanabe}, {Weinberg}, {Yanny},
    {Yasuda}, \& {SDSS Collaboration}}]{York.Adelman.ea00} {York},
  D.~G., {Adelman}, J., {Anderson}, Jr., J.~E., {et~al.} 2000, \aj,
  120, 1579

\end{thebibliography}
\end{document}